\documentclass{aa}
\usepackage{float}
\usepackage{graphicx}
\usepackage{txfonts}
\usepackage{amsmath}
\usepackage{siunitx}
\usepackage[colorlinks=true,linkcolor=blue,citecolor=blue,urlcolor=blue]{hyperref}
\usepackage{orcidlink}
\usepackage{stfloats}

\graphicspath{{figures/}}

\newcommand{\feh}{[\mathrm{Fe/H}]}
\newcommand{\mh}{[\mathrm{M/H}]}
\providecommand{\nodata}{\ensuremath{\cdots}}

\begin{document}

\title{Estimating stellar metallicities from Gaia DR3 XP data using LAMOST DR10}
\titlerunning{Estimating stellar metallicities from Gaia DR3 XP using LAMOST DR10}
\authorrunning{Srivastava et al.}

\author{D. Srivastava\,\orcidlink{0009-0009-1630-4643}\inst{1}
           \and A. Niedzielski\,\orcidlink{0000-0002-0587-8854}\inst{1}
           \and R. Smiljanic\,\orcidlink{0000-0003-0942-7855}\inst{2}}

\institute{Institute of Astronomy, Nicolaus Copernicus University in Toru\'n, ul. Gagarina 11, 87-100 Toru\'n, Poland\\
   \email{divyansh@doktorant.umk.pl, Andrzej.Niedzielski@umk.pl}
    \and
    Nicolaus Copernicus Astronomical Center, Polish Academy of Sciences, ul. Bartycka 18, 00-716 Warsaw, Poland}

\date{Received; Accepted}

\abstract
{\emph{Gaia} DR3 provides astrophysical parameters for hundreds of millions of stars but the [M/H] from its GSP-Phot module suffer from systematic bias.}
{In this paper, we estimate stellar metallicities from \emph{Gaia} DR3 data, using the homogeneous spectroscopic iron abundances $\feh$ of LAMOST DR10 as training labels.}
{We have cross-matched LAMOST DR10 $\feh$ with \emph{Gaia} DR3 and trained a gradient-boosted decision-tree regressor (XGBoost) on \num{1.20} million AFGK stars, using only \emph{Gaia}-derived inputs and proxies. We validated the estimates on held-out LAMOST stars, on GALAH DR4 and APOGEE DR17, and on \num{46} open clusters, and applied the model to measure the radial metallicity gradient of the disk of the Milky Way.}
{On the held-out test set, the model reaches a mean absolute error of \SI{0.052}{dex} and $R^2=0.94$ with negligible bias, against \SI{0.242}{dex} for GSP-Phot on the same stars. The estimates transfer well to external surveys with a mean absolute error of \SI{0.066}{dex} (GALAH) and \SI{0.068}{dex} (APOGEE). For open clusters, the median difference between our estimated $\feh$ and $\feh$ from spectroscopy surveys is \SI{0.041}{dex} which is smaller than both GSP-Phot (\SI{0.248}{dex}) and previous work on an APOGEE-trained XGBoost model (\SI{0.067}{dex}). When we applied our model to the disk, it recovers a broken thin disk radial gradient (inner $+0.119$, outer $-0.058\,\mathrm{dex\,kpc^{-1}}$ with a break near \SI{5.9}{kpc}) and an open-cluster gradient of $-0.066\,\mathrm{dex\,kpc^{-1}}$, both in agreement with previous high-resolution spectroscopy works.}
{Our estimated $\feh$ values are accurate to \SIrange{0.05}{0.07}{dex} for AFGK stars within the range $\feh\gtrsim-2.5$; below this limit the predictions should be treated as lower bounds. The catalogue and the trained model are publicly available on Zenodo. These estimates are suitable for chemical studies of the Milky Way.}


\maketitle
\nolinenumbers

\section{Introduction}
\label{sec:intro}

The metallicity of a star is very important for Galactic archaeology. Together with kinematics and ages, the metallicity helps us trace how the Milky Way formed, enriched, and mixed its stellar populations \citep{FreemanBland2002}. The \emph{Gaia} mission \citep{GaiaMission2016} has transformed this field by providing astrometry and spectrophotometry for more than a billion stars. But at this scale, turning low-resolution data into accurate abundances is very difficult.

\emph{Gaia}~DR3 \citep{GaiaDR3} publishes astrophysical parameters from the Apsis (astrophysical parameter inference system) pipeline \citep{Fouesneau2023}. The GSP-Phot module estimates temperature, gravity, extinction, and metallicity from the low-resolution blue and red photometer (BP/RP, hereafter ``XP'') spectra, apparent G magnitude, and parallax \citep{Creevey2023, AndraeGSPPhot2023}. The XP spectra are released as coefficients of a continuous basis-function representation \citep{DeAngeli2023}. While GSP-Phot performs well for bulk stellar parameters, its metallicities are known to suffer from systematic trends with colour, magnitude, and extinction \citep{AndraeGSPPhot2023}.

A productive response has been to re-derive XP-based metallicities empirically, anchored to large spectroscopic surveys. \citet{Andrae2023} trained an XGBoost regressor on APOGEE labels, supplemented by very metal-poor stars from \citet{Li2022}. Using the XP coefficients, its derived products, and CatWISE photometry, they produced metallicity estimates for 175 million stars that are more accurate than GSP-Phot. \citet{Zhang2023} used a forward-modelling approach. They used LAMOST DR8 labels to model the XP spectra together with 2MASS and WISE photometry. These works established that the XP coefficients contain a usable metallicity signal once a clean spectroscopic training set anchors it.

In this work, we estimate stellar metallicities from \emph{Gaia} DR3 data, trained on LAMOST DR10 labels. LAMOST \citep{Cui2012, Zhao2012} is among the largest optical spectroscopic surveys available. Its DR10 low-resolution AFGK catalogue provides homogeneous stellar parameters, including $\feh$, for several million stars through the LAMOST Stellar Parameter Pipeline (LASP). Compared with APOGEE, LAMOST offers a larger training set in the optical, where the XP spectra are the most informative. It also provides better coverage of the metal-poor tail without the need of artificial augmentation from an external survey. LAMOST also includes many stars in the dusty parts of the disk near the Galactic plane, where dust makes the stars redder and makes it harder to measure metallicity from the low-resolution XP spectra. APOGEE also covers these regions, but with fewer stars and fewer types of stars.  We learn a mapping from \emph{Gaia}-only inputs to the LAMOST $\feh$ scale with a gradient-boosted decision-tree model (XGBoost) \citep{Chen2016}. XGBoost is chosen for its accuracy on heterogeneous tabular data, its efficiency at survey scale, and its transparent feature diagnostics.

We then test our estimates with three independent tests: external validation against GALAH DR4 \citep{Buder2025} and APOGEE DR17 \citep{Abdurrouf2022}; validation against \num{46} open clusters with high-resolution metallicities; and a measurement of the radial metallicity gradient of the disk. Throughout the paper, we compare our results with the raw GSP-Phot metallicities on identical stars, providing a like-for-like measure of the improvement.

The paper is organised as follows. Section~\ref{sec:data} describes the data, cross-match, and selection. Section~\ref{sec:methods} describes the model, training, and uncertainty estimation. Section~\ref{sec:results} reports the held-out performance and diagnostics, and Section~\ref{sec:external} the external validation. Section~\ref{sec:gradient} applies our model to the disk gradient, and Section~\ref{sec:clusters} to open clusters. Section~\ref{sec:discussion} discusses limitations and usage, and Section~\ref{sec:conclusions} concludes.

\section{Data}
\label{sec:data}

\subsection{Spectroscopic labels: LAMOST DR10}
\label{sec:lamost}
The iron abundances $\feh$ from the LAMOST DR10 low-resolution AFGK stellar-parameter catalogue are our training labels. It is derived by the LAMOST Stellar Parameter Pipeline (LASP). We curated the catalogue to make sure that only high-quality labels are used for training of the model.

We first reject the observations flagged by the pipeline as fibre offsets to keep only the genuine stellar targets. We then apply $\mathrm{S/N}$ thresholds in three bands, $\mathrm{S/N}_g\ge30$, $\mathrm{S/N}_r\ge30$, and $\mathrm{S/N}_i\ge20$. This is stricter than the survey's nominal limits so that the spectra carry enough signal for a reliable parameter fit. Additionally, we select objects with well-determined atmospheric parameters, with formal errors $\sigma_{\rm [Fe/H]}\le\SI{0.10}{dex}$, $\sigma_{T_{\rm eff}}\le\SI{120}{K}$, $\sigma_{\log g}\le\SI{0.20}{dex}$, and $\sigma_{\rm RV}\le\SI{10}{km\,s^{-1}}$. A star whose temperature or gravity is poorly constrained is unlikely to have a trustworthy $\feh$, so our cuts remove ambiguous fits even though we use only $\feh$ for the label. Finally, we restrict the labels to $-2.5\le\feh\le1.0$. The DR10 LRS stellar-parameter catalogue contains \num{7450303} entries, of which \num{3846576} spectra survive these quality cuts. The resulting label distribution is concentrated near solar metallicity: the median $\feh$ is $-0.13$ and the 1st--99th percentile range is $[-1.04, +0.39]$. Metal-poor stars make up only about \SI{1}{\percent} of the sample, but in absolute terms this is still around \num{13000} stars with $\feh<-1$ which is enough to train on without augmenting the sample from an external survey. However, their small fractional weight leads the model to underestimate the most metal-poor stars (see Section~\ref{sec:external}).

Many stars have more than one LAMOST spectrum. For each \emph{Gaia} source we combine its repeat measurements into a single inverse variance weighted mean,\begin{equation}
\feh = \frac{\sum_k w_k\,\feh_k}{\sum_k w_k}, \qquad w_k = \sigma_{\mathrm{[Fe/H]},k}^{-2}
\label{eq:labelmean}
\end{equation}

We clip per-measurement error to $[0.03,0.30]$~dex so that no single spectrum dominates. A label uncertainty $\sigma_{\rm label}=(\sum_k w_k)^{-1/2}$ is also calculated. If repeat measurements for a source has scatter more than \SI{0.20}{dex}, then it is discarded. The combined label set contains \num{2810415} unique stars. These inverse variance weights are used only to construct the per--star label and are not used in the training loss itself. The weights used in the training loss are the conditional colour--magnitude weights that we introduce in Section~\ref{sec:cmdw}.

\subsection{\emph{Gaia} DR3 inputs}
\label{sec:gaia}

The input vector that we feed to the model has \num{139} features. All inputs are \emph{Gaia} DR3 catalogue quantities or simple combinations of them, grouped as follows.

\paragraph{XP spectral coefficients (110)} The core inputs are the \num{55} BP and \num{55} RP coefficients of the continuous XP representation. These coefficients encode the shape of the low-resolution spectrum and carry information about the metallicity. The raw coefficients scale with apparent flux, so two stars of identical spectral shape but different brightness have very different coefficient amplitudes. We remove this dependence by rescaling each star's coefficients to the flux level of a $G=15$ source, following \citet{Andrae2023}, so that the model sees spectral \emph{shape} and not the brightness.

\begin{equation}
c_i' = c_i \times 10^{\,(G-15)/2.5},
\label{eq:xpnorm}
\end{equation}

\paragraph{Broad-band photometry and colours} We use three mean magnitudes $G$, $G_{\rm BP}$, $G_{\rm RP}$ and the colours $G_{\rm BP}-G_{\rm RP}$ and $G-G_{\rm RP}$ to summarise the overall spectral energy distribution (SED) and constrain temperature.

\paragraph{Astrometry and quality indicators} We include parallax $\varpi$ with its error and $\mathrm{S/N}$ $\varpi/\sigma_\varpi$; proper motions $\mu_{\alpha^*},\mu_\delta$; Renormalised Unit Weight Error (RUWE) for which high values can be taken as an indicator of binarity, crowding, or calibration issues; the BP/RP flux-excess factor; the five-parameter astrometric uncertainty; Gaia's synthetic magnitude in the RVS (Radial Velocity Spectrometer) band ($G_{\rm RVS}$); and radial velocity with its error. On top of flagging data quality, these parameters carry information about distance and kinematics.

\paragraph{Luminosity and kinematic proxies} The other most useful supporting information for breaking the temperature--gravity--metallicity degeneracy is luminosity. At fixed temperature, dwarfs and giants differ in gravity, and in the field, they separate cleanly in their absolute magnitude. Therefore, we make parallax-based absolute-magnitude proxies in each band, similar in spirit to the parallax-based pseudo-absolute magnitudes of \citet{Andrae2023} (who adopt a form linear in parallax, whereas we use the logarithmic form),

\begin{equation}
M_X^{\rm proxy} = X + 5\log_{10}(\varpi/1000) + 5, \qquad X\in\{G,\,G_{\rm BP},\,G_{\rm RP}\},
\label{eq:absmag}
\end{equation}
with $\varpi$ in mas and no extinction correction at this stage. For stars with noisy parallaxes we also add the reduced proper motion which serves as a kinematic luminosity proxy,
\begin{equation}
H_G = G + 5\log_{10}\mu + 5, \qquad \mu=\sqrt{\mu_{\alpha^*}^2+\mu_\delta^2}\ \mathrm{[mas\,yr^{-1}]},
\label{eq:rpm}
\end{equation}

We also include the parallax$\times$magnitude cross-terms like $\varpi G$, $\varpi G_{\rm BP}$, $\varpi G_{\rm RP}$ which encodes the distance and apparent brightness together and helps separate the populations more sharply than either quantity alone.

\paragraph{Extinction terms} Interstellar reddening shifts the colours and the SED. This would otherwise be mistaken for a change in temperature or metallicity. So we supply the GSP-Phot line-of-sight dust estimates such as the $G$-band extinction $A_G$ and the reddening $E(G_{\rm BP}-G_{\rm RP})$ together with the dereddened colour $(G_{\rm BP}-G_{\rm RP})_0=(G_{\rm BP}-G_{\rm RP})-E(G_{\rm BP}-G_{\rm RP})$, an extinction-corrected absolute-magnitude proxy $M_{G,0}^{\rm proxy}=M_G^{\rm proxy}-A_G$, and the total-to-selective ratio $A_G/|E(G_{\rm BP}-G_{\rm RP})|$ as a proxy for the local dust law. These are the only GSP-Phot products we use, and they describe the intervening \emph{dust}, not the star i.e. they are physically independent of $\feh$. We deliberately do not use the GSP-Phot atmospheric parameters ($T_{\rm eff}$, $\log g$, and especially [M/H]) because we found them to inject a metallicity prior into the predictions. We keep them only as independent diagnostics.

\paragraph{Galactic latitude} Lastly, we include Galactic latitude $b$ as the only sky-position input. It gives the model a weak information about the extinction column and about the changing thin/thick disk mix with height above the plane. We verify in Section~\ref{sec:heldout} that the test residuals are flat with $b$ (Figure ~\ref{fig:resid_lat_plx}). So the feature does not introduce any spurious spatial structure, and its importance is modest (rank 22 of \num{139} by gain); we revisit this choice in Section~\ref{sec:discussion}.

\subsection{Cross-match and selection}
\label{sec:xmatch}

After the feature selection, we cross-match the cleaned LAMOST label set with \emph{Gaia} DR3 by \texttt{source\_id} and join the XP coefficient table. On the \emph{Gaia} side we then make a filter for positive parallax, colour $0.2\le G_{\rm BP}-G_{\rm RP}\le1.9$, BP/RP flux-excess factor between \num{1.0} and \num{1.30}, and $\mathrm{RUWE}<1.40$. The positive-parallax requirement is imposed because the absolute-magnitude proxies of Equation~(\ref{eq:absmag}) involve $\log_{10}\varpi$ and are undefined otherwise. For this bright cross-matched sample it removes only \num{2037} stars (\SI{0.09}{\percent} of the stars passing all other cuts). We do not put any explicit temperature cut because the colour range should ideally set the effective temperature interval of the sample. We drop stars missing any of the above required inputs. After all the cuts, the resulting working sample contains \num{1197937} stars. A significant fraction of this sample lies in extincted regions: \SI{18}{\percent} of the stars have GSP-Phot extinction $A_G>0.5$~mag.

\begin{figure*}[!b]
\centering
\includegraphics[width=\textwidth]{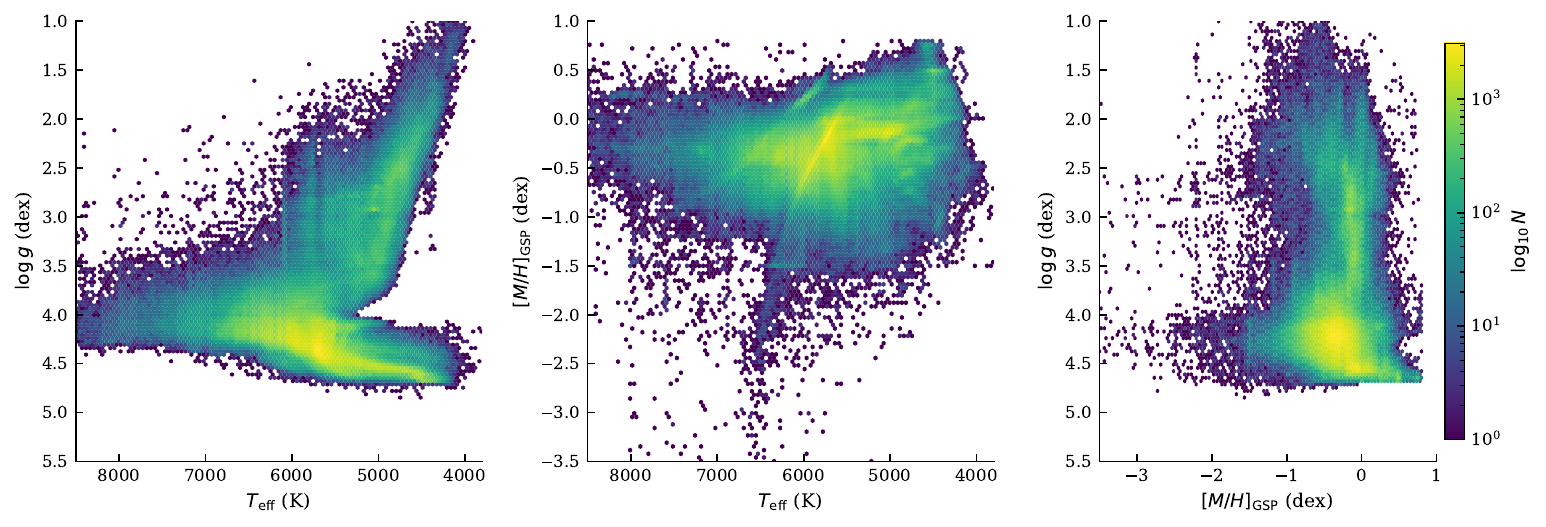}
\caption{Density of the working sample in the $(T_{\rm eff},\log g)$, $(T_{\rm eff},\mh)$, and $(\mh,\log g)$ pairs. The atmospheric parameters shown are the \emph{Gaia} GSP-Phot values which are not model inputs. The sample is selected on colour ($0.2\le G_{\rm BP}-G_{\rm RP}\le1.9$), not on temperature; the axes are clipped to the populated range. About \SI{0.4}{\percent} of stars have GSP-Phot $T_{\rm eff}>\SI{8500}{K}$ (they are the warm stars near the blue colour limit for which the GSP-Phot temperature is unreliable) and are outside the displayed range.}
\label{fig:sample}
\end{figure*}

\begin{figure*}[!h]
\centering
\includegraphics[width=0.85\textwidth]{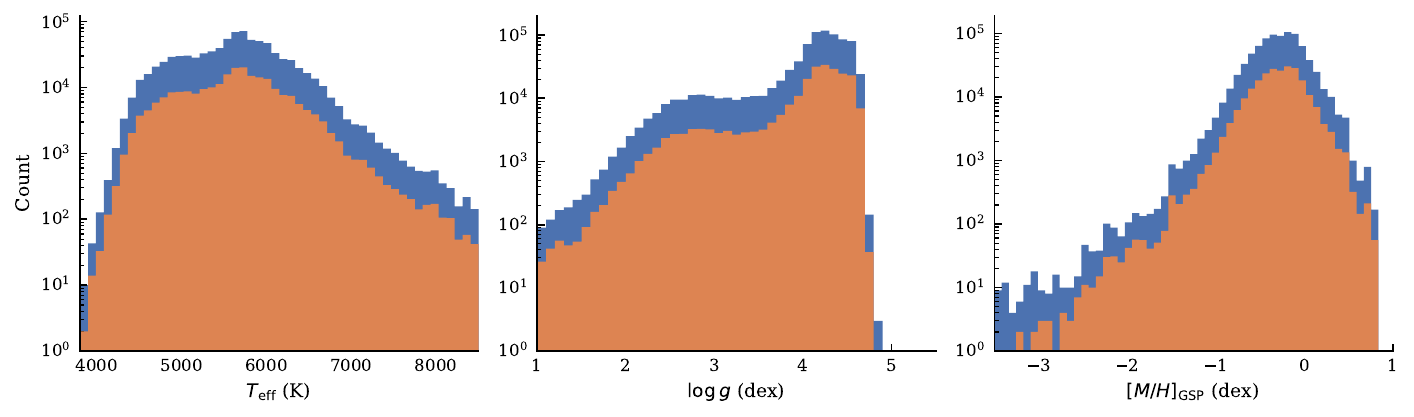}
\caption{Distributions of \emph{Gaia} GSP-Phot effective temperature, surface gravity, and metallicity for the training (blue) and held-out test (orange) sets, on logarithmic count scales. It confirms that the test set is representative of the training distribution.}
\label{fig:paramdist}
\end{figure*}

We split the sample $70/10/20$ into training (\num{838555}), validation (\num{119794}), and test (\num{239588}) sets. We stratify it in quantile bins of $\feh$ and colour so that every region of parameter space is represented in each subset. The model fitting uses the training set only; the validation set is kept for early stopping; the test set is used only for final reporting. Figure~\ref{fig:sample} shows the temperature--gravity--metallicity distribution of the sample and Figure~\ref{fig:paramdist} shows the distribution of these three GSP-Phot values in the training and held-out test sets.

\section{Method}
\label{sec:methods}

\subsection{Model Training}
\label{sec:model}
We use the gradient-boosted decision tree model (XGBoost) \citep{Chen2016} to model $\feh$. XGBoost is well suited to tabular problems with heterogeneous feature scales and has been used effectively for stellar label inference such as \citet{Andrae2023}. For the training, we use the squared-error objective. Up to \num{2000} boosting rounds are allowed (each round adds one more decision tree that tries to fix the errors the current model is making). The maximum tree depth is \num{8}. Maximum tree depth decides how many split layers a single tree can grow. Ideally deeper layers mean more flexibility but it is also more prone to memorise noise. The learning rate is \num{0.04} (a shrinkage factor on each new tree's contribution; a small value means each tree adds only a little, so the model improves gradually and is less likely to overshoot). Row and column subsampling are \num{0.8} and \num{0.7} i.e. every tree is trained on a random \SI{80}{\percent} of the stars and \SI{70}{\percent} of the features which helps all the trees to stop looking alike and reduces overfitting. The minimum child weight is \num{20} (a leaf must contain at least that much total weight before the model is allowed to split it further. This is done to prevent the model from carving out tiny groups that fit noise). L2 and L1 leaf regularisation are $\lambda=1.0$ and $\alpha=0.1$ (they are the penalty terms on the size of leaf values that pull the predictions toward zero and smooth the model). Early stopping monitors the validation mean absolute error every round, with a patience of \num{50} i.e. if it does not improve for \num{50} consecutive rounds, training will stop and not continue till round \num{2000}.

\subsection{Weakening the colour-conditioned prior}
\label{sec:cmdw}

Since our training sample is comprised of magnitude limited field stars, a model trained on it will not just learn the signal related to metallicity but also the field correlation between the colour and metallicity that comes from the mix of Galactic population. So, when the model later sees any chemically homogeneous system such as an open cluster, where every star really has the same metallicity but spans a wide range of colours, it gives them different predicted metallicities depending on their position in the colour--magnitude diagram (CMD). This appears as a spurious trend of predicted $\feh$ with colour (which we mention as the ``hump'' in Section ~\ref{sec:hump}).

To weaken this learned prior, we apply a sample weighting during training. If we weight stars so that, inside any small region of the CMD, the model sees just as many metal-rich stars as metal-poor stars, then ideally, it should not learn any relation between position of the star in CMD and its metallicity. We do this weighting on the dereddened colour--magnitude plane $\bigl((G_{\rm BP}-G_{\rm RP})_0,\,M_{G,0}^{\rm proxy}\bigr)$. We use the dereddened colour and absolute magnitude so that dust along the line of sight doesn't smear stars into the wrong bins. We divide the plane into 16 colour bins $\times$ 16 magnitude bins (hereon "cell"), each with the same number of stars. Inside each CMD cell, stars are further divided into 10 equal population sub-cells of $\feh$. Each training star gets a weight,

\begin{equation}
w \;=\; \biggl(\frac{1}{N_{\rm cell}}\biggr)^{\!0.35}\,\cdot\,\frac{N_{\rm cell}}{N_{\rm cell,[Fe/H]}},
\label{eq:cmdw}
\end{equation}

where $N_{\rm cell}$ is the population of its CMD cell and $N_{\rm cell,[Fe/H]}$ that of its CMD$\,\times\,$[Fe/H] sub-cell. The first factor upweights sparse CMD cells mildly. The 0.35 exponent softens its influence so that rare CMD corners are not allowed to dominate the fit. In the second factor, by weighting each star by $N_{\rm cell}/N_{\rm cell,[Fe/H]}$, an under-populated sub-cell gets large per-star weights and an over-populated one gets small per-star weights, in exact proportion. This removes the colour--metallicity coupling that the model would have otherwise reproduced. The combined weight is normalised to unit mean and clipped to [0.20, 7.0] to bound the contribution of any single star. This weighting is the only term carried into the training loss as a per-star weight. Although it does not remove the hump in the cluster predictions of $\feh$ (Section~\ref{sec:hump}), it reduces it at a very low cost to field accuracy.

\subsection{Predictive uncertainties}
\label{sec:uncertainty}
The $\feh$ prediction reported throughout this paper is from the single primary model described above. To obtain a per-star uncertainty we also train a five-member ensemble, i.e. five extra XGBoost models, each fit on an independent random resample of the training stars drawn from the same training set. The five members use the same architecture, features, and per-star CMD weights as the primary model. The per-star uncertainty is the standard deviation across the five ensemble predictions for that star. This uncertainty is model variance rather than a formal noise estimate. It inflates for stars in sparse regions of feature space and for atypical inputs. We note that noise on the input features (e.g. the XP coefficient, photometric, and parallax uncertainties) also translates into scatter in the predictions. The median ensemble scatter on the held-out test set is \SI{0.013}{dex}.

\section{Results}
\label{sec:results}

\subsection{Comparison with GSP-Phot}
\label{sec:heldout}
\begin{figure}[H]
\centering
\includegraphics[width=\columnwidth]{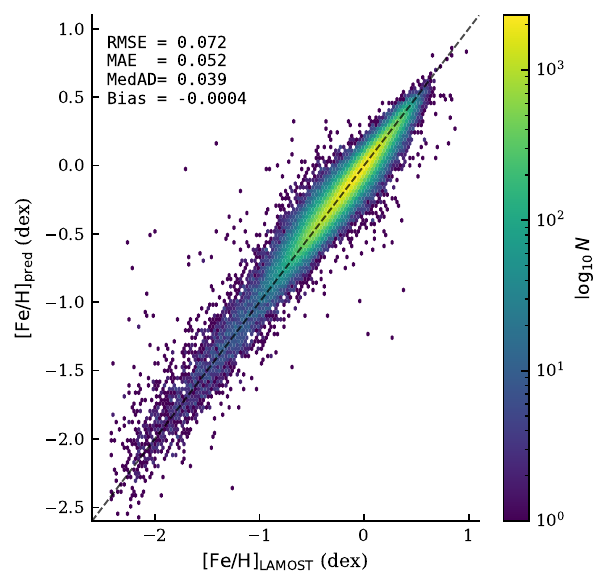}
\caption{Predicted versus spectroscopic $\feh$ on the held-out LAMOST test set. The dashed line is the one-to-one relation.}
\label{fig:predtrue}
\end{figure}

We note that all results in this paper come from the single primary model trained on the fixed 70/10/20 split of Section~\ref{sec:xmatch} and we did not perform k-fold cross-validation. On the held-out test with \num{239588} stars, the model recovers the LAMOST $\feh$ scale with RMSE \SI{0.072}{dex}, MAE \SI{0.052}{dex}, a median absolute deviation (MedAD) of \SI{0.039}{dex}, a bias of \SI{-0.0004}{dex}, and $R^2=0.94$ (Figure~\ref{fig:predtrue}). For the same stars, the published \emph{Gaia} DR3 GSP-Phot metallicities give RMSE $=\SI{0.31}{dex}$, MAE $=\SI{0.24}{dex}$, an offset of \SI{-0.14}{dex}, and $R^2<0$. This is roughly five times worse on either bulk metric. However the raw GSP-Phot product is not the strongest available baseline. A head-to-head comparison against the XP-based metallicity estimates of \citet{Andrae2023} is presented in Section~\ref{sec:cluster_acc}.

\begin{figure*}[h]
\centering
\includegraphics[width=\textwidth]{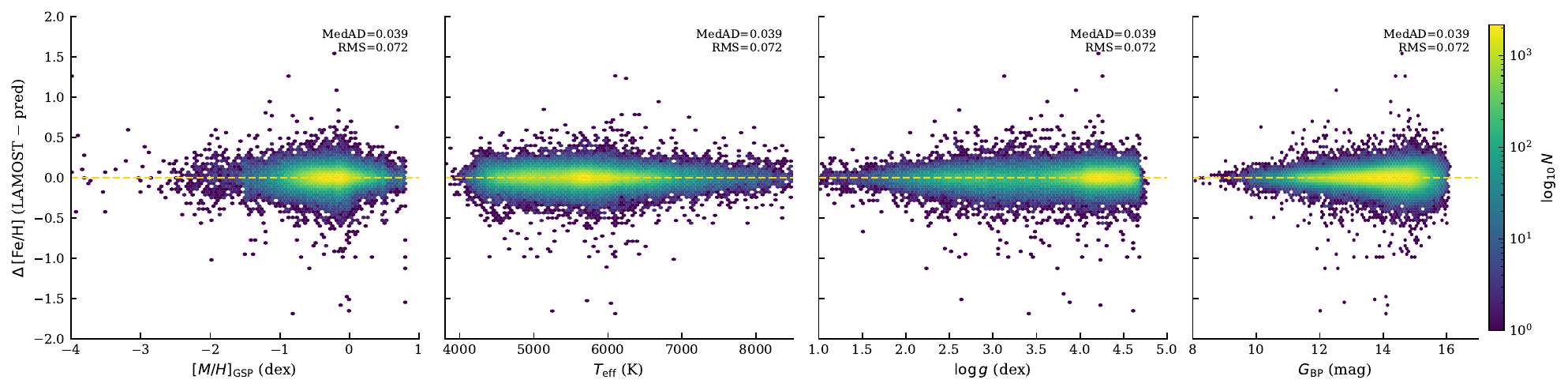}
\caption{Test-set residuals $\Delta=\feh_{\rm LAMOST}-\feh_{\rm pred}$ versus the GSP-Phot metallicity, effective temperature, surface gravity, and $G_{\rm BP}$. The dashed line marks zero; each panel lists the median absolute deviation and RMS of $\Delta$.}
\label{fig:resid_params}
\end{figure*}

In figure~\ref{fig:resid_params}, we show the test residuals $\Delta=\feh_{\rm LAMOST}-\feh_{\rm pred}$ against the GSP-Phot metallicity, effective temperature, surface gravity, and $G_{\rm BP}$. The residual field is centred near zero and tight across the AFGK range and shows no strong trend in any projection. 

\begin{figure}[H]
\centering
\includegraphics[width=0.85\columnwidth]{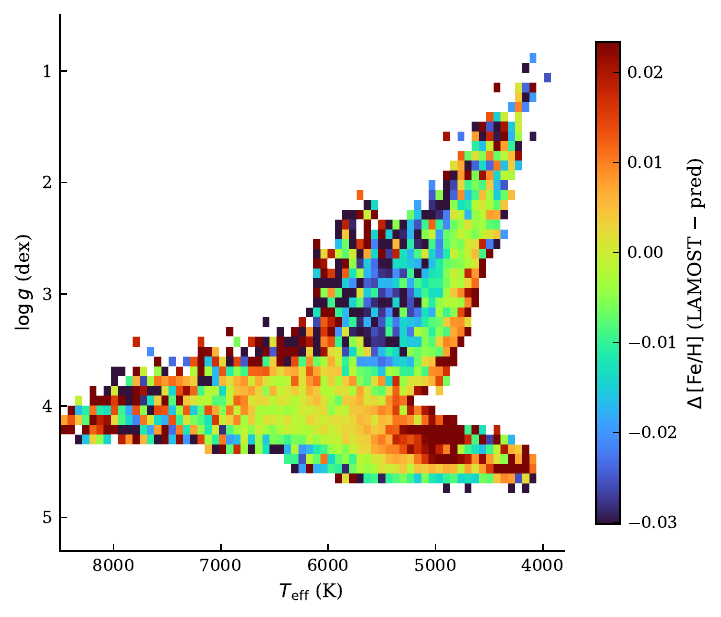}
\caption{Binned mean test-set residual $\Delta=\feh_{\rm LAMOST}-\feh_{\rm pred}$ across the $T_{\rm eff}$--$\log g$ plane. Bins with fewer than five stars are blank and the colour scale shows only the 10th--90th percentile of the binned residual.}
\label{fig:resid_hr}
\end{figure}

\begin{figure}[H]
\centering
\includegraphics[width=0.85\columnwidth]{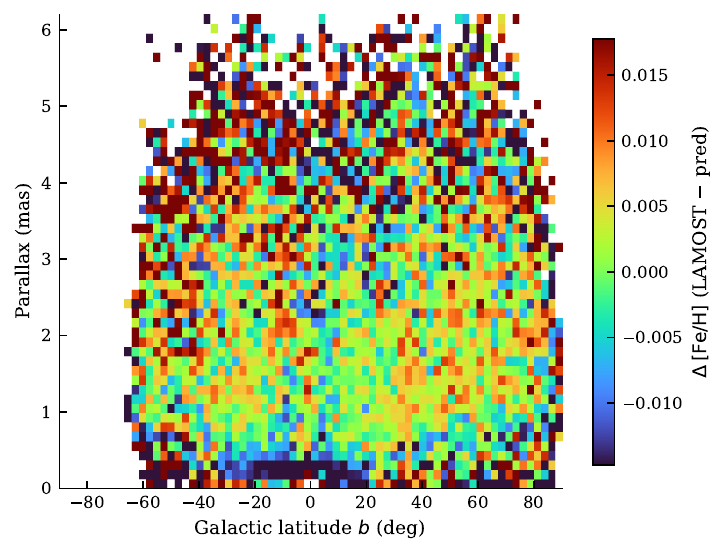}
\caption{Binned mean test-set residual $\Delta=\feh_{\rm LAMOST}-\feh_{\rm pred}$ as a function of Galactic latitude and parallax. Bins with fewer than five stars are blank and the colour scale shows only the 10th--90th percentile of the binned residuals.}
\label{fig:resid_lat_plx}
\end{figure}

In figure~\ref{fig:resid_hr}, we map the residuals across the HR diagram, confirming that $\feh$ from the model for dwarfs and giants are recovered without any luminosity-class distinction, and Figure~\ref{fig:resid_lat_plx} shows that the residuals are flat with Galactic latitude and parallax. The latter is a good check on our inclusion of $b$ as a feature (Section~\ref{sec:gaia}). The absence of latitude structure in the residuals confirms that the model is not exploiting $b$ to encode any sky-dependent bias.

We note again that the selection is on colour rather than temperature (Section~\ref{sec:xmatch}). A small fraction (\SI{0.4}{\percent}) of test stars have GSP-Phot temperatures above \SI{8500}{K} near the blue colour limit of the sample. These are retained in all reported metrics; the residual figures display the \SIrange{3800}{8500}{K} interval that contains \SI{99.6}{\percent} of the sample.

\subsection{Feature importance}
\label{sec:featimp}
\begin{figure}
\centering
\includegraphics[width=\columnwidth]{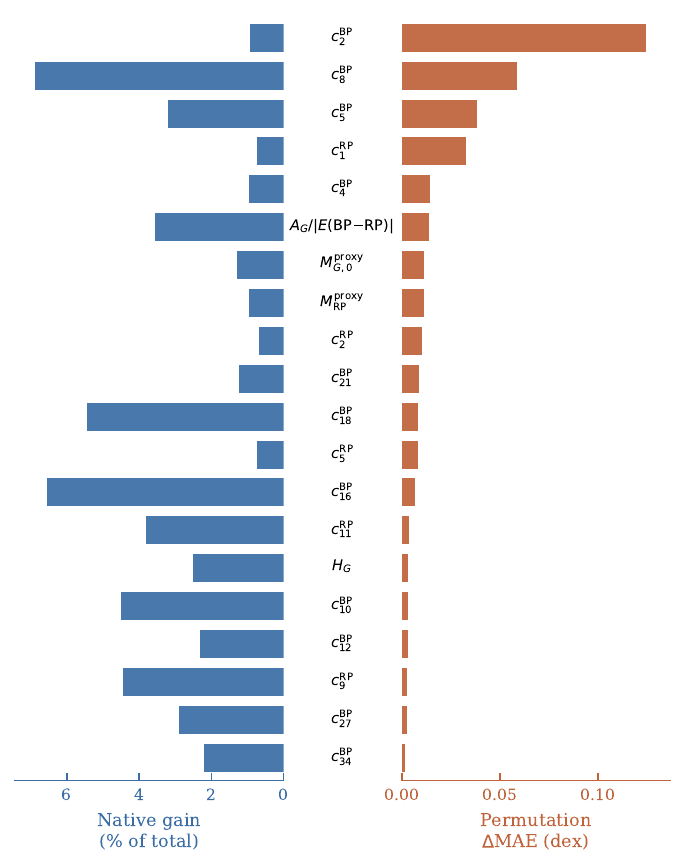}
\caption{Feature importance for the trained model, ordered vertically by descending permutation importance. \emph{Left wing} (blue): native XGBoost gain, the average loss reduction per split accumulated over training, as a percentage of the total gain over all features. \emph{Right wing} (orange): permutation $\Delta$MAE on the held-out test set, i.e.\ the increase in mean absolute error (in dex) when each feature's values are randomly shuffled.}
\label{fig:featimp}
\end{figure}

To study the importance of the features, we used two standard metrics, gain and permutation. Gain increases every time the trees split on a feature during training. Each split adds to that feature's gain budget. The more useful a feature is for the trees to split on, the larger its budget grows. On the other hand, permutation takes the trained model and shuffles a feature's value at test time. If that feature's information is genuinely unique then shuffling hurts the permutation. We found that most of the model's predictive power lies in the low-order BP coefficients $c^{\rm BP}_{1\text{--}8}$ (Fig.~\ref{fig:featimp}). Permuting $c^{\rm BP}_2$ alone raised the test MAE by $\sim\SI{0.13}{dex}$, roughly twice as much as any other feature. The low-order coefficients encode the broadband slope of the BP spectrum, the spectral region where iron-peak line blanketing depresses the blue flux most strongly \citep{AndraeGSPPhot2023}. So the model is showing exactly the band where the metallicity signal should live. The higher-order BP coefficients ($c^{\rm BP}_{16}$, $c^{\rm BP}_{18}$, $c^{\rm RP}_9$) are split on heavily by the trees but contribute little when permuted. This means that the model uses them as redundant copies of the low-order signal.

Among non-XP inputs, only four enter the top twenty. These are the dust ratio $A_G/|E(\mathrm{BP}-\mathrm{RP})|$, the dereddened absolute-magnitude proxy $M_{G,0}^{\rm proxy}$, its red-band sibling $M_{\rm RP}^{\rm proxy}$, and the reduced proper motion $H_G$. We included these features specifically to break the temperature--gravity--metallicity degeneracy (Section ~\ref{sec:gaia}).

\subsection{External validation: GALAH DR4 and APOGEE DR17}
\label{sec:external}
To test our trained model to different independent scales, we apply it, without any retuning, to stars in GALAH DR4 \citep{Buder2025} and APOGEE DR17 \citep{Abdurrouf2022}.
For the validation, we removed all stars that appeared in the LAMOST training, validation, or test sets.

\begin{figure*}[!h]
\centering
\includegraphics[width=0.8\textwidth]{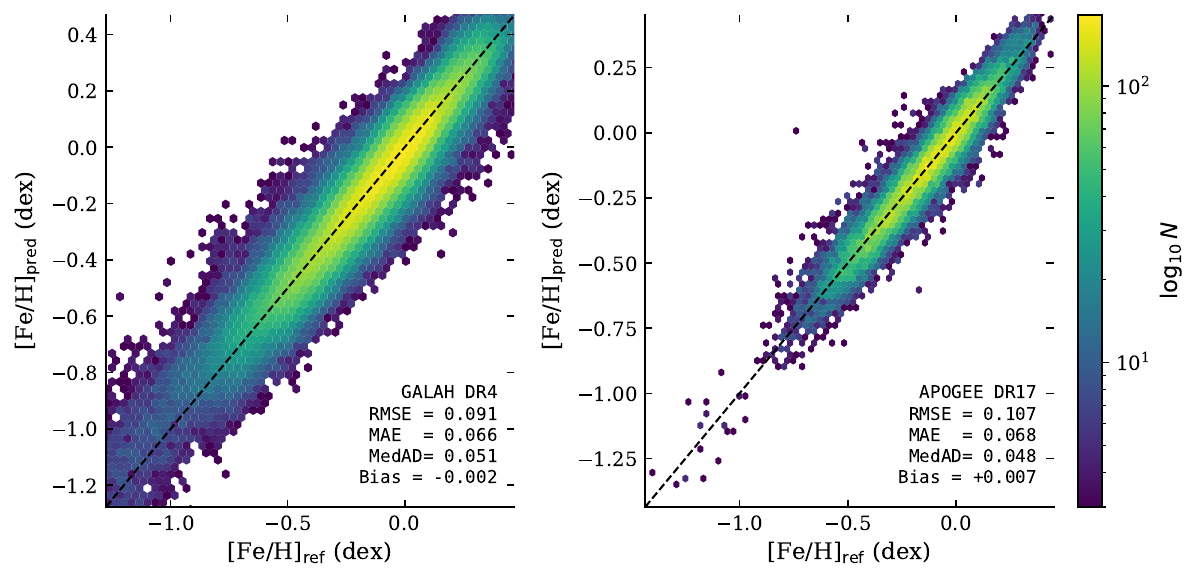}
\caption{External validation against GALAH DR4 (left) and APOGEE DR17 (right): predicted $\feh$ versus the reference spectroscopic $\feh$. Each panel lists RMSE, MAE, MedAD, and bias for that survey. The dashed line is the one-to-one relation.}
\label{fig:external_scatter}
\end{figure*}

Figure~\ref{fig:external_scatter} compares the predictions with the reference $\feh$. Against GALAH we obtain RMSE $=\SI{0.091}{dex}$, MAE $=\SI{0.066}{dex}$, MedAD $=\SI{0.051}{dex}$, and a bias of \SI{-0.002}{dex} ($R^2=0.91$). Against APOGEE, RMSE $=\SI{0.107}{dex}$, MAE $=\SI{0.068}{dex}$, MedAD $=\SI{0.048}{dex}$, and a bias of \SI{+0.007}{dex} ($R^2=0.88$). These errors are only slightly larger than on the held-out LAMOST set. Comparisons in literature find LAMOST--APOGEE scatters of $\sim\SI{0.13}{dex}$ in broad samples \citep{Anguiano2018}, down to a MedAD of $\sim\SI{0.04}{dex}$ under stricter quality cuts \citep{Soubiran2022}.

\begin{figure*}
\centering
\includegraphics[width=0.9\textwidth]{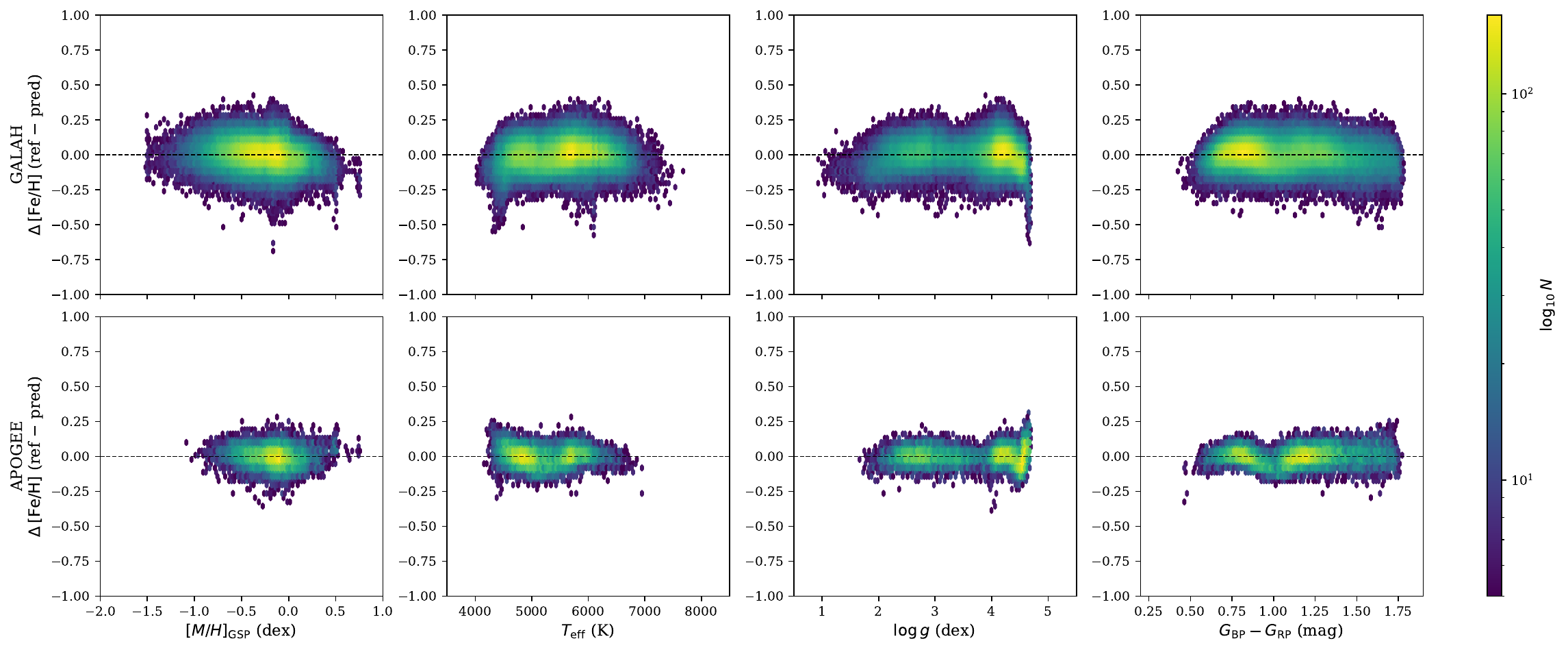}
\caption{External residuals $\Delta=\feh_{\rm ref}-\feh_{\rm pred}$ versus the GSP-Phot metallicity, effective temperature, surface gravity, and $G_{\rm BP}-G_{\rm RP}$ colour, for GALAH DR4 (top) and APOGEE DR17 (bottom).}
\label{fig:external_resid}
\end{figure*}

If we quadrature-subtract our LAMOST-internal RMSE of \SI{0.072}{dex} from the external RMSEs, it gives scale gaps of \SI{0.055}{dex} for GALAH and \SI{0.079}{dex} for APOGEE. The APOGEE value sits comfortably within the documented LAMOST--APOGEE scatter, and the GALAH value is smaller still. So the model is mostly preserving the LAMOST scale rather than imposing its own systematics. In figure~\ref{fig:external_resid}, we show the external residuals against the GSP-Phot metallicity, effective temperature, surface gravity, and BP-RP colour. They are mostly flat to first order in every projection, with noticeable deviation for $\log g > 4$ for both GALAH and APOGEE. And against BP-RP the behaviour is more interesting: against GALAH it is essentially flat (hump metric $\approx-\SI{0.01}{dex}$), while against APOGEE there is a $\sim\!\SI{0.05}{dex}$ dip at BP-RP$\,\approx\,1$ same colour range as the cluster hump of Section~\ref{sec:hump} though with smaller amplitude (\SI{70}{\percent} of the cluster value). We do not attempt to characterise the survey-to-survey difference in this work. It is worth keeping in mind that no spectroscopic survey is free of systematics. For example, \citet{Jonsson2020} discuss this issue for APOGEE DR16 and show that some level of systematic pattern as a function of $\log g$ remains in their abundances even after calibration (see their Fig.~13). Another such pattern appears in the work of \citet{Wheeler2020}, where the authors train a parameter inference system on GALAH DR2 results and inherit systematic problems from the survey (see their Fig.~11). Among the reasons for such issues are that there is no single pipeline that can work for every type of star \citep[see e.g.][]{Smiljanic2014} and that 3D non-LTE effects are usually not well understood and not taken into account \citep[see][for a review]{Lind2024}.

\begin{figure}
\centering
\includegraphics[width=\columnwidth]{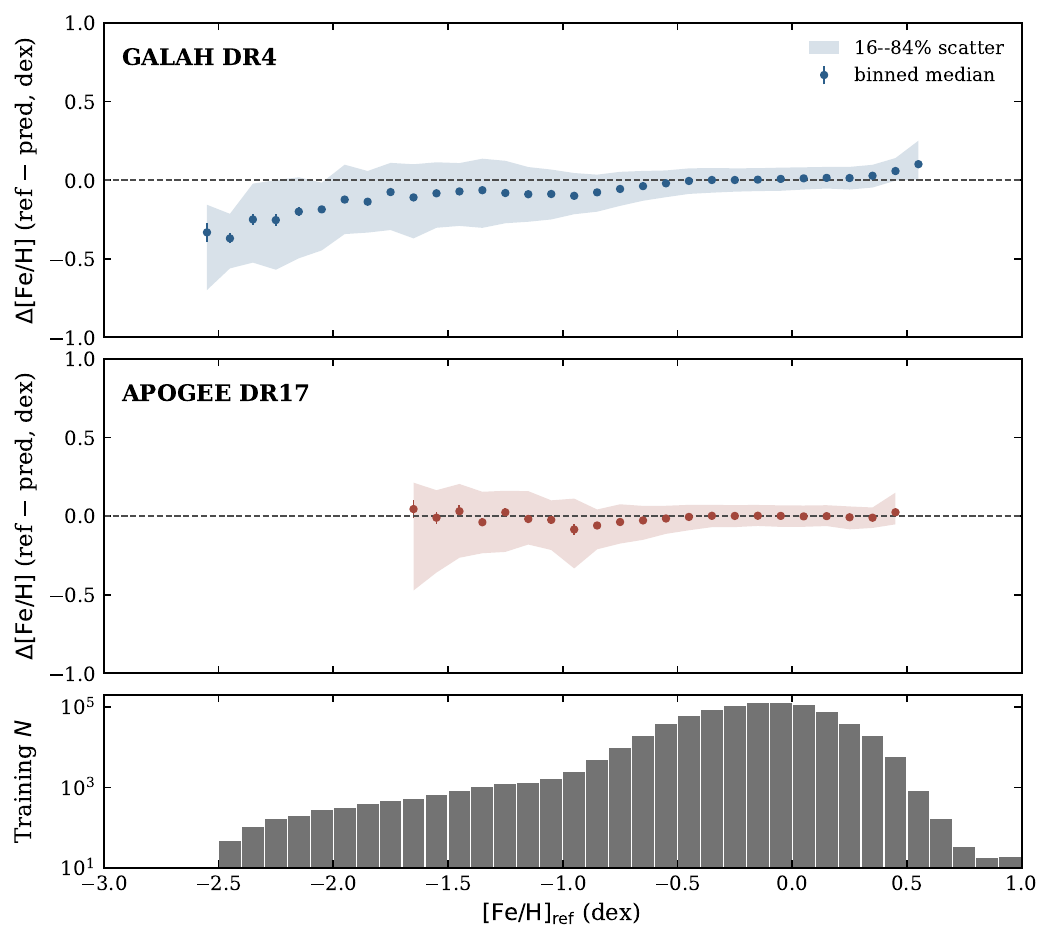}
\caption{Binned median residual $\Delta=\feh_{\rm ref}-\feh_{\rm pred}$ against reference $\feh$ for GALAH DR4 (top) and APOGEE DR17 (middle), in 0.1 dex bins. Points show the binned median and the shaded bands show the 16th--84th percentile of $\Delta$ within each bin. The bottom panel shows the LAMOST training $\feh$ histogram on the same axis. We do not show the bins with less than 10 stars.}
\label{fig:bias_external}
\end{figure}

But beside the global bias numbers that we report above, it is important to note separately, the structure at the metal-poor tail of the GALAH's $\feh$ distribution. Figure~\ref{fig:bias_external} shows the binned median residual against the reference $\feh$ for both surveys, together with the LAMOST training stars. It is evident that around the training peak, both surveys agree: across $-1\le\feh\le+0.5$ the binned median bias stays below \SI{0.10}{dex} against GALAH and \SI{0.085}{dex} against APOGEE. The two surveys diverge at the metal-poor side. Against GALAH, a clear shrinkage can be already seen at $\feh\!\lesssim\!-1.0$, growing to $\sim\!0.2$--$\SI{0.4}{dex}$ by $\feh=-2.0$ and exceeding \SI{1}{dex} at the extreme tail. On the other hand, against APOGEE, the bias remains within $\sim\SI{0.1}{dex}$ down to the limit of the APOGEE sample at $\feh\!\sim\!-2.3$. We do not try a full reconciliation of the difference, which may show the very different luminosity class composition of the two surveys in the metal-poor regime (APOGEE giants vs GALAH dwarfs). The shrinkage seen against GALAH is the regression-to-the-mean behaviour expected of any squared-error regressor where the training density gets sparse. Although it is not an error of the well-sampled peak, it does set a practical lower limit on how metal-poor a star our model can identify in absolute terms. So, the very metal-poor tail obtained from our trained model should be treated as lower bounds on $\feh$ rather than precise abundances.

\section{Application: the radial metallicity gradient of the disk}
\label{sec:gradient}
We apply the model to a random sample of \num{4850766} LAMOST-less \emph{Gaia} DR3 stars. We use the same Gaia-side quality cuts as the training set (Section~\ref{sec:xmatch}) with not-null values for most of the fundamental training features except a few. Values such as radial velocity and the related $G_{\rm RVS}$ magnitude are missing for $\sim$\SI{89}{\percent} of these sources. It is because the RVS spectrograph publishes velocities only down to $G_{\rm RVS}=14$~mag and is moreover restricted to FGK templates ($T_{\rm eff}\in[3100,6750]$~K) below $G_{\rm RVS}=12$~mag \citep{Katz2023}. Imposing not-null values for the RVS quantities would have been against the random selection. XGBoost routes these missing values through its default-branch behaviour, so the RVS quantities are used where available and skipped where not. Our random field sample has no spectroscopic labels, so we cannot quantify how much the absence of the RVS features affects the accuracy of the predictions. However, we can measure how much the predictions themselves change. To do this, we repeated the prediction for the \num{530299} catalogue stars that do have RVS information, this time removing their radial velocity, its uncertainty, and $G_{\rm RVS}$. The effect is small: the predictions without the RVS features are more metal-poor by \SI{0.02}{dex} (median shift), with an RMS difference of \SI{0.033}{dex}; \SI{89}{\percent} of the stars change by less than \SI{0.05}{dex} and only \SI{0.8}{\percent} by more than \SI{0.1}{dex}. The impact is largest in the sparsely populated metal-poor tail ($\feh<-1$), where the RMS difference reaches \SI{0.086}{dex}. These differences are below the field accuracy of the model (Section~\ref{sec:external}), so the missing RVS features degrade but do not fundamentally change the predictions. Nevertheless, the small systematic offset should be kept in mind when mixing stars with and without RVS data. The predictions for the stars without the RVS features carry a larger per-star ensemble uncertainty than those for stars with RVS features. We also note that the kinematic disk-membership selection used in Section~\ref{sec:bensby} requires finite RVS features, so the gradient measurements below are unaffected by this.

\begin{figure*}
\centering
\includegraphics[width=0.85\textwidth]{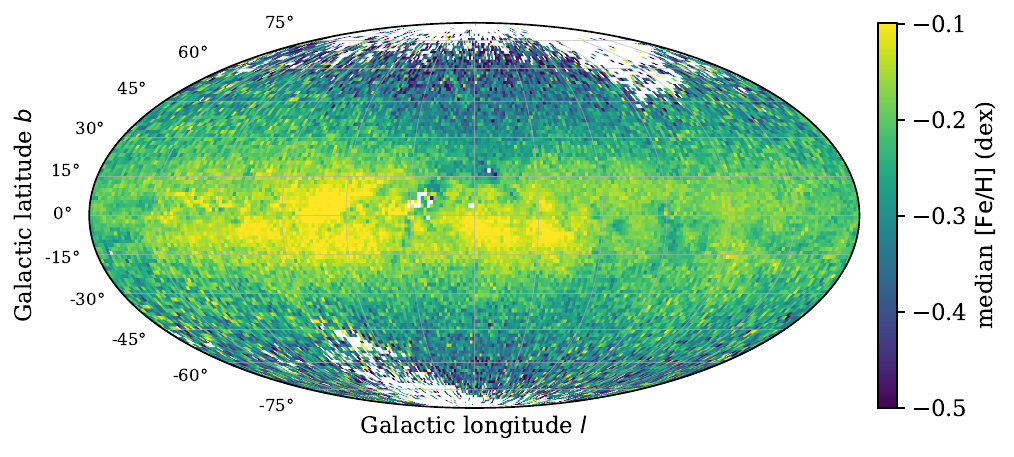}
\caption{All-sky median predicted [Fe/H] in Galactic coordinates for the \num{2076814} catalogue stars with $\varpi>\SI{0.5}{mas}$ ($d<\SI{2}{kpc}$). Each cell is the median over its members and cells with fewer than five stars are blank. Colour bar is $[2,98]$ percentile clipped.}
\label{fig:skymap}
\end{figure*}

Figure~\ref{fig:skymap} shows the resulting all-sky median [Fe/H]. We restrict the sample to \num{2076814} stars with $\varpi>\SI{0.5}{mas}$ ($d<\SI{2}{kpc}$), where the parallax distances are precise and the brighter XP spectra provide the most reliable metallicities. The metal-rich disk plane and the more metal-poor high-latitude populations are immediately visible. A linear regression of our model's predicted [Fe/H] on $|z|$ over $0.3<|z|<\SI{1.6}{kpc}$ has a slope of $-0.237$~dex\,kpc$^{-1}$ which matches the SEGUE G-dwarf value of $-0.243^{+0.039}_{-0.053}$~dex\,kpc$^{-1}$ measured over the same height range \citep{Schlesinger2014}. And if we restrict our sample to G-dwarf-like colours only ($G_{\rm BP}-G_{\rm RP}\in[0.8,1.1]$), it gives us a steeper $-0.248$~dex\,kpc$^{-1}$, in even closer agreement with the G-dwarf-only Schlesinger sample. Also, extending our fit to the wider $|z|<\SI{2}{kpc}$ range of \citet{Nandakumar2017} gives $-0.204$~dex\,kpc$^{-1}$ which is consistent with their multi-survey combined value of $-0.241\pm0.028$~dex\,kpc$^{-1}$ within $\sim\!1\sigma$ of the quoted uncertainty.

From the agreement of these vertical gradient predictions, we proceed to measure the radial metallicity gradient in both thin and thick disk, a quantity for which many independent spectroscopic studies exist and against which our almost photometrically trained model can be judged.

\subsection{Disk membership and geometry}
\label{sec:bensby}

Instead of a simple $|z|$ cut, we separate the thin and thick disks kinematically. The two populations have heavily overlapping vertical density profiles (scale heights of $\sim\!300$ and $\sim\!900$~pc; \citealt{Juric2008,BlandHawthorn2016}), so any geometric selection will unavoidably mix them near the plane. In contrast, the Bensby velocity ellipsoids are well separated in 3D and provide much cleaner samples \citep{Bensby2003}. We use their local velocity dispersions and density fractions. For each star, we calculate the relative probabilities $\mathrm{D}$, $\mathrm{TD}$, and $\mathrm{H}$ of being drawn from the thin disk, thick  disk, and halo velocity ellipsoids weighted by their local space-density fractions ($0.94$, $0.06$, and $0.0015$), and use the ratios $\mathrm{TD/D}$ and $\mathrm{TD/H}$ as membership indicators. We calculate the velocities from \emph{Gaia} positions, proper motions, and radial velocities, with distances $d=1/\varpi$, and correct it to the local standard of rest with the solar motion of \citet{Schonrich2010}. Finite radial velocity, finite proper motions, and a parallax $\mathrm{S/N}$ above \num{10} is needed to limit the fractional distance errors to below \SI{10}{\percent}. Galactocentric coordinates use $R_0=\SI{8.2}{kpc}$ \citep{BlandHawthorn2016} and $Z_0=\SI{0.025}{kpc}$. Stars with $\mathrm{TD/D}<0.1$ are assigned to the thin disk (\num{358531} stars) and those with $\mathrm{TD/D}>10$ and $\mathrm{TD/H}>1$ to the thick disk (\num{35035} stars). We exclude the ambiguous intermediate cases and the \num{4816} halo stars ($\mathrm{TD/H}<1$) from the further fits because our purpose of keeping the halo component in the classification was to keep the halo interlopers out of the thick disk sample.

\begin{figure}
\centering
\includegraphics[width=\columnwidth]{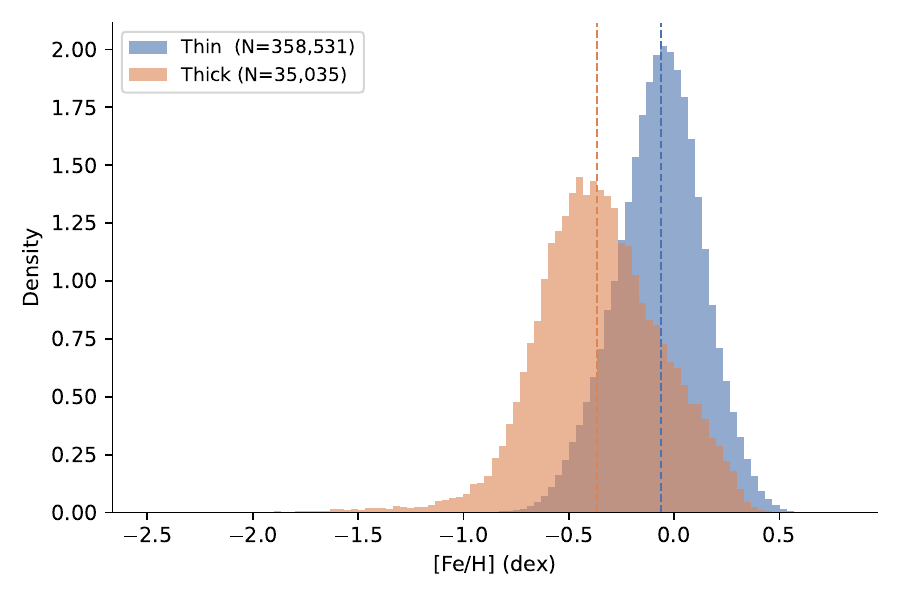}
\caption{Metallicity distribution functions of the kinematically selected thin and thick disk samples, drawn from the \num{486182} stars in the \num{4.85} million-star catalogue that satisfy the Bensby selection. Dashed lines mark the medians ($-0.06$ and $-0.37$~dex). The thick disk is correctly more metal poor than the thin disk.}
\label{fig:mdf}
\end{figure}

Figure~\ref{fig:mdf} shows the two kinematically separated disk populations. The thin disk has a median $\feh=-0.06$ and the thick disk $-0.37$. As a kinematic sanity check we compute the velocity dispersions of the two samples, i.e. the standard deviations of the Galactic velocity components $(U,V,W)$ measured relative to the local standard of rest (LSR). The thin disk sample has $(\sigma_U,\sigma_V,\sigma_W)=(37,20,15)\,\mathrm{km\,s^{-1}}$, matching the thin-disk velocity ellipsoid of \citet{Bensby2003}, $(35,20,16)\,\mathrm{km\,s^{-1}}$. The thick disk sample has larger dispersions, $(103,43,54)\,\mathrm{km\,s^{-1}}$, above the Bensby thick-disk ellipsoid of $(67,38,35)\,\mathrm{km\,s^{-1}}$, because the $\mathrm{TD/D}>10$ selection keeps only the stars most confidently assigned to the thick disk, i.e. its high-velocity tail.

\subsection{Gradients}
\label{sec:grad_results}
\begin{figure}
\centering
\includegraphics[width=\columnwidth]{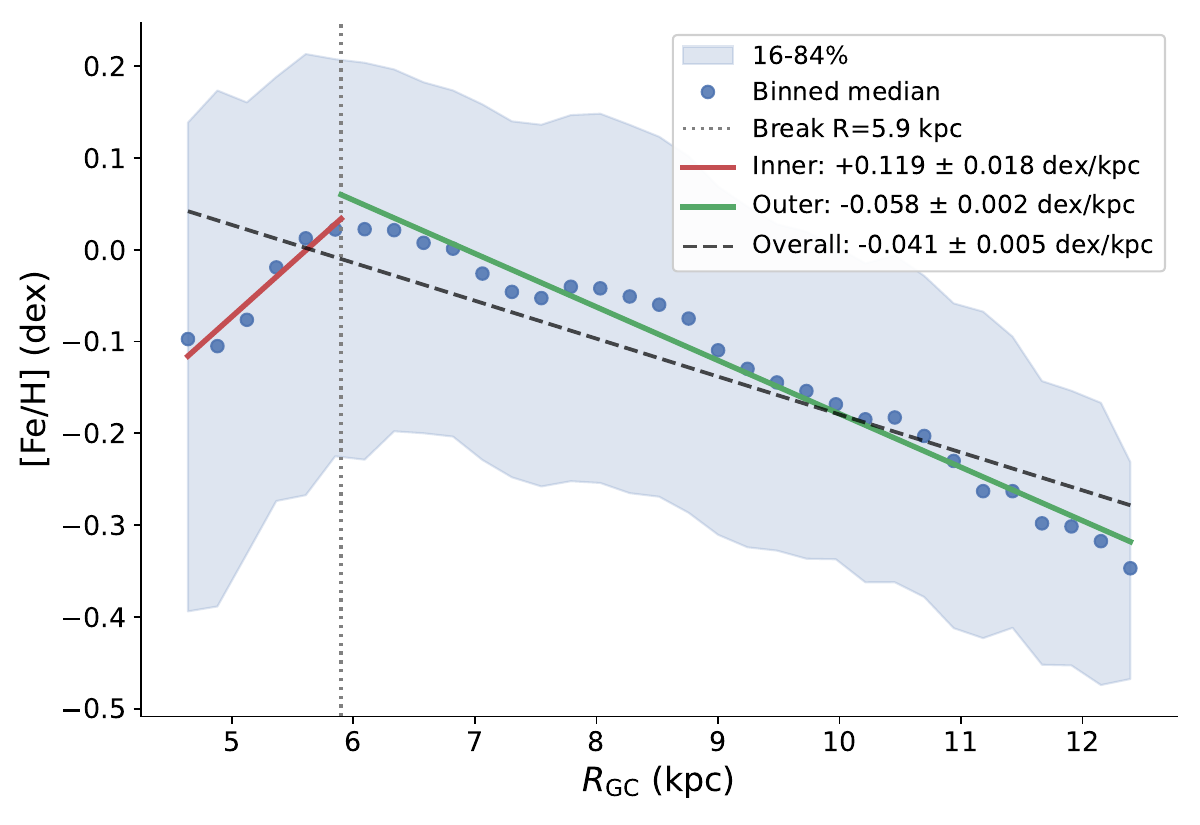}
\caption{Thin disk radial metallicity profile from this work for the kinematically selected Bensby sample. The profile shows the expected inner-rising, outer-declining slopes.}
\label{fig:gradient}
\end{figure}

We fit binned median profiles of [Fe/H] versus galactocentric radius. We note that the uncertainties reported throughout this subsection are the 1$\sigma$ equivalent spread of slope values across 500 randomised fits of the binned-median profile. Bins with less than \num{200} stars are dropped, so the profile begins at the first populated bin near $R\approx\SI{4.6}{kpc}$. Before fitting, we comment briefly on the survey selection function. The kinematic selection of Section~\ref{sec:bensby} requires a finite \emph{Gaia} radial velocity, which effectively limits the sample to $G_{\mathrm{RVS}}\lesssim14$~mag. At these bright magnitudes \emph{Gaia} is essentially complete over the volume probed and the selection does not depend on metallicity at fixed position. Gradient studies with deeper samples would need to model the selection function explicitly but for the bright sample used here we expect its influence on the measured slopes to be small. For the thin disk, a single linear fit over the populated range $4.6<R<\SI{12.4}{kpc}$ gives a slope of $-0.041\pm0.006\,\mathrm{dex\,kpc^{-1}}$. However, the profile is better understood if we treat it as broken at $R\approx\SI{5.9}{kpc}$. Then the inner disk has a \emph{positive} slope of $+0.119\pm0.018\,\mathrm{dex\,kpc^{-1}}$ and the outer disk a negative slope of $-0.058\pm0.002\,\mathrm{dex\,kpc^{-1}}$ (Figure~\ref{fig:gradient}, left). This shape is qualitatively the same as the spectroscopic determinations of \citet{Lian2023} where they find a break at a later $6.9\pm0.6$~kpc with inner and outer slopes of $+0.031\pm0.010$ and $-0.052\pm0.008\,\mathrm{dex\,kpc^{-1}}$. Our outer slope agrees with theirs and with the overall Cepheid slope value of $-0.060\pm0.002\,\mathrm{dex\,kpc^{-1}}$ from \citet{Genovali2014}. It also agrees with the \emph{Gaia} DR3 chemical-cartography analysis of \citet{GaiaRecioBlanco2023}, who find a radial metallicity gradient of $-0.056\pm0.007\,\mathrm{dex\,kpc^{-1}}$ for $R>\SI{7}{kpc}$ close to the Galactic plane from the RVS-based GSP-Spec metallicities of giant stars. One should note that GSP-Spec is derived from the medium-resolution RVS spectra and not from the XP spectra, so this agreement is an independent cross-check between the two \emph{Gaia} metallicity channels. Our break radius is somewhat smaller than the \citeauthor{Lian2023} value. Their 1$\sigma$ lower bound is \SI{6.3}{kpc}, marginally above our upper bound and our inner slope is steeper. But we do not attempt to attribute the inner-slope difference to a single cause here. We will note that the two measurements probe physically different quantities. \citet{Lian2023} report a light-weighted integrated profile from SDSS spectra whereas we report binned medians of individual, kinematically selected thin disk stars. The thick disk shows a very shallow radial gradient ($-0.007\pm0.004\,\mathrm{dex\,kpc^{-1}}$) as expected for a radially mixed old population \citep{Cheng2012} and consistent with the near-uniformity of the high-[$\alpha$/Fe] metallicity distribution across the disk reported by \citet{Hayden2015}.


\section{Validation on open clusters}
\label{sec:clusters}

Open clusters are chemically homogeneous to a few hundredths of a dex \citep{Bovy2016}. More importantly, they lie in regions of the colour--magnitude diagram and distance distribution that differ from the field which makes them a strict, training-independent test. We use the membership of \num{46} nearby clusters from \citet{GaiaHRD2018} and apply our trained model to each member star in the same way as for the field stars. Cluster distances are the inverse of the median member parallax. We cross-match these clusters to the compilation of \citet{Spina2022} to get \num{26} clusters with reference [Fe/H] and list them in Table~\ref{tab:clusters}.

\subsection{Comparison with other catalogues}
\label{sec:cluster_acc}

\begin{figure}
\centering
\includegraphics[width=0.82\columnwidth]{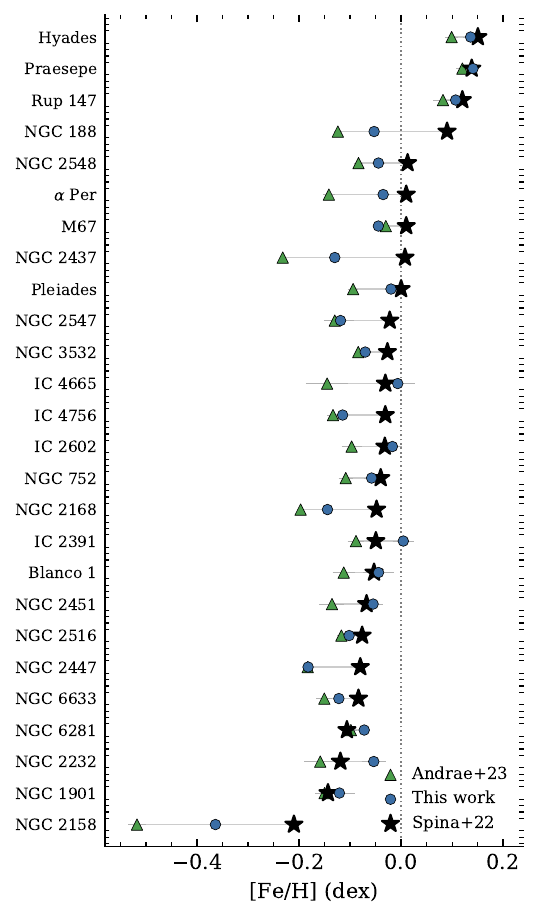}
\caption{Per-cluster metallicity for the \num{26} clusters in common, ordered by the \citet{Spina2022} value. Our estimates sit systematically closer to the \citet{Spina2022}.}
\label{fig:cluster_compare}
\end{figure}

The cluster-median metallicities reproduce the high-resolution values with a MedAD of \SI{0.041}{dex} (mean \SI{0.053}{dex}). This is comparable to the held-out LAMOST test accuracy of Section~\ref{sec:results}. The five nearest benchmark clusters are recovered to within \SI{0.02}{dex}, including the Hyades ($+0.136$ vs.\ $+0.150$) and Praesepe ($+0.141$ vs.\ $+0.138$). Figure~\ref{fig:cluster_compare} compares our values with those of \citet{Spina2022} and with the APOGEE-trained model of \citet{Andrae2023}, which we cross-matched to the same members. Our MedAD (\SI{0.041}{dex}) is smaller than that of \citet[][\SI{0.067}{dex}]{Andrae2023}. The GSP-Phot median for the same clusters is \SI{0.248}{dex}. Therefore our LAMOST-trained model reproduces the spectroscopic cluster $\feh$ at least as well as the APOGEE-trained model and far better than the GSP-Phot $\mh$.

\subsection{The intra-cluster colour trend}
\label{sec:hump}

\begin{figure}
\centering
\includegraphics[width=\columnwidth]{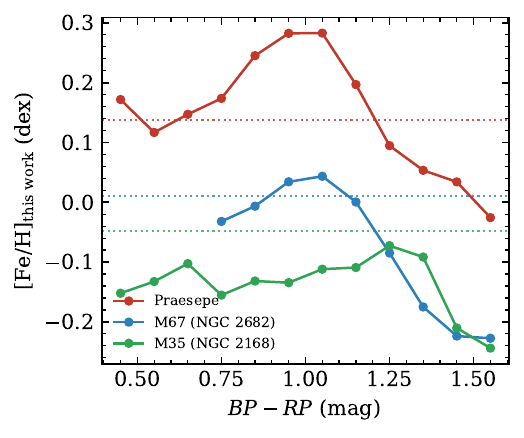}
\caption{Binned median predicted [Fe/H] versus $G_{\rm BP}-G_{\rm RP}$ for Praesepe, M67, and M35. Dotted horizontal lines are the \citet{Spina2022} value of each cluster.}
\label{fig:hump_profile}
\end{figure}

We found that in many clusters the predicted $\feh$ is not constant with colour. It rises to a maximum near $G_{\rm BP}-G_{\rm RP}\approx1.0$ and declines toward redder colours, even though the true [Fe/H] should be fixed for a cluster. We quantify this central ``hump'' as the difference between the median prediction in a central colour window $G_{\rm BP}-G_{\rm RP}\in[0.90,1.10]$ and in two flanking windows $[0.55,0.75]$ and $[1.25,1.45]$ combined (Table~\ref{tab:clusters}). Median amplitude of the hump across the sample is \SI{0.07}{dex}. Importantly, the amplitude of the hump changes notably from cluster to cluster, and not only because of colour sampling or distance to the clusters. Figure~\ref{fig:hump_profile} compares three nearby and equally well-populated clusters that all span the full colour range. The hump is strong in Praesepe (metal-rich), moderate in M67 (near-solar), and effectively \emph{absent} in M35 (NGC\,2168), whose profile stays flat through the central colours despite having several hundred members in each window. Therefore the hump is not a fixed feature and is a heavily cluster-dependent systematic.

We performed a controlled experiment in which we trained an adversarial two-branch network with a nuisance and spectroscopic branch. We provided the nuisance branch all the photometric and colour--magnitude features and penalised it for carrying any metallicity information. This drove that branch's output to $\sim\!10^{-6}$~dex. However, the hump still showed up in the spectroscopic branch which was trained only on the XP coefficients. The hump is therefore encoded in the XP spectra themselves. At the resolution of the XP data ($R\sim50$) the coefficients usually carry the broadband spectral shape which is set by the temperature. In our LAMOST field training set where the temperature correlates with [Fe/H] through the Galactic age--metallicity relation, the model learns a temperature-conditioned metallicity prior. And when we apply our trained model to a chemically homogeneous cluster, where member stars have the same [Fe/H] but on a wide temperature range, this learned prior projects onto colour as the hump. We did not find any correlation of the amplitude of the hump with distance, age, or metallicity. Through the conditional weighting of Section~\ref{sec:cmdw}, we try to reduce it but we could not remove it. The same hump is also visible in the APOGEE field residuals at about \SI{70}{\percent} of the cluster amplitude (Section~\ref{sec:external}), though it is not present in the GALAH sample. The same hump is also present in the APOGEE-trained model of \citet{Andrae2023}. Our median amplitude of the hump (\SI{0.069}{dex}) is in fact larger than theirs (\SI{0.038}{dex}) but on absolute accuracy of cluster's $\feh$ predictions, we are the more accurate of the two (Section~\ref{sec:cluster_acc}).

We also considered if the hump could be an artefact of crowding. In dense fields, the dispersed XP spectra of neighbouring sources overlap on the CCD and extracted spectra become unreliable regardless of the inference method. \citet[][their Fig.~27]{Creevey2023} show this in the globular cluster $\omega$~Cen where the GSP-Phot metallicity estimates degrade strongly toward the crowded centre. They report that the BP/RP windows begin to overlap at a source density of about \num{600000} per square degree ($\approx\num{170}$ per arcmin$^2$). To check whether our clusters fall in this regime, we counted the \emph{Gaia} DR3 sources within \SI{5}{arcmin} of each cluster centre. Only two of our clusters reach this density, and both lie at low Galactic latitude where the counts are dominated by unrelated foreground and background stars rather than cluster members. More importantly, the hump does not behave as one would expect from the crowding effect. Its amplitude does not increase with field density and the strongest humps occur in the sparsest fields like Praesepe, the Pleiades and NGC~1039, all below \num{7} sources~arcmin$^{-2}$ (one to two orders of magnitude below the overlap threshold). The hump is strong in Praesepe ($\sim$\num{2} sources~arcmin$^{-2}$) yet almost absent in the ten-times denser M35. Therefore, we conclude that overlapping XP spectra do not drive the hump.

\subsection{The cluster radial gradient}
\label{sec:cluster_grad}

\begin{figure}
\centering
\includegraphics[width=\columnwidth]{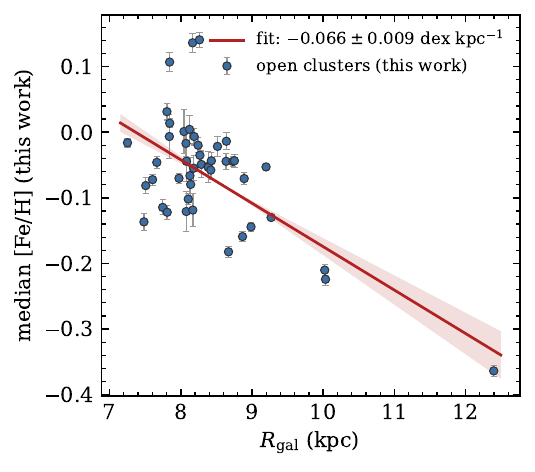}
\caption{Cluster median [Fe/H] from this work versus cylindrical galactocentric radius for all \num{46} clusters. Error bars are the standard error of each median; the red line is a weighted linear fit with its $1\sigma$ band.}
\label{fig:cluster_grad}
\end{figure}

Independent of the field's stellar analysis of metallicity gradient (Section~\ref{sec:gradient}), the cluster $\feh$ are also as a standard measure. We computed each cluster's cylindrical galactocentric radius from its mean line of sight and parallactic distance (with $R_0=\SI{8.2}{kpc}$, \citealt{BlandHawthorn2016}). A weighted linear fit of cluster-median [Fe/H] versus $R_{\rm GC}$ over $\SIrange{7.3}{12.4}{kpc}$ gives a slope of \SI{-0.066\pm0.009}{dex\per kpc} (Figure~\ref{fig:cluster_grad}). This matches the open-cluster gradient $-0.068 \pm0.001\,\mathrm{dex\,kpc^{-1}}$, measured from APOGEE by \citet{Donor2020} and $-0.066 \pm 0.007\,\mathrm{dex\,kpc^{-1}}$ from compiled samples \citep{Netopil2016}. Although, in our calculations, the slope is set by a few clusters beyond \SI{9}{kpc}. Our predictions for outer clusters are systematically more metal-poor than the \citet{Spina2022} measurements by $\sim\!\SI{0.14}{dex}$ while the inner clusters are nearly unbiased (Table~\ref{tab:clusters}). Therefore, our measured slope should be taken as the upper bound on the true gradient.

\begin{table*}
\centering
\caption{Open clusters with spectroscopic metallicities from \citet{Spina2022}, ordered by
heliocentric distance. Distances are the inverse of the median member parallax. $|\Delta_{\rm hump}|$ is the
intra-cluster colour-trend amplitude (Sect.~\ref{sec:hump}); \nodata\ marks clusters with too few
members across the $BP-RP$ range to measure it.}
\label{tab:clusters}
\begin{tabular}{lrrcccc}
\hline\hline
Cluster & $N$ & $d$ (pc) & $\mathrm{[Fe/H]_{this\,work}}$ & $\mathrm{[Fe/H]_{Spina+22}}$ & $\Delta$ & $|\Delta_{\rm hump}|$ \\
\hline
Hyades & 81 & 47 & $+0.136\pm0.015$ & $+0.150\pm0.063$ & $-0.014$ & \nodata \\
Pleiades & 193 & 135 & $-0.020\pm0.012$ & $+0.000\pm0.050$ & $-0.020$ & 0.166 \\
IC\,2391 & 45 & 150 & $+0.004\pm0.021$ & $-0.050\pm0.020$ & $+0.054$ & \nodata \\
IC\,2602 & 50 & 150 & $-0.017\pm0.019$ & $-0.032\pm0.136$ & $+0.015$ & \nodata \\
$\alpha$\,Per & 125 & 174 & $-0.035\pm0.013$ & $+0.010\pm0.050$ & $-0.045$ & 0.107 \\
Praesepe & 201 & 185 & $+0.141\pm0.011$ & $+0.138\pm0.041$ & $+0.003$ & 0.187 \\
NGC\,2451 & 38 & 191 & $-0.055\pm0.019$ & $-0.068\pm0.019$ & $+0.013$ & \nodata \\
Blanco\,1 & 34 & 235 & $-0.044\pm0.031$ & $-0.053\pm0.048$ & $+0.009$ & \nodata \\
NGC\,6774 (Rup\,147) & 52 & 307 & $+0.107\pm0.014$ & $+0.120\pm0.030$ & $-0.013$ & \nodata \\
NGC\,2232 & 21 & 323 & $-0.054\pm0.024$ & $-0.119\pm0.037$ & $+0.065$ & \nodata \\
IC\,4665 & 28 & 345 & $-0.007\pm0.033$ & $-0.031\pm0.078$ & $+0.024$ & \nodata \\
NGC\,2547 & 47 & 386 & $-0.119\pm0.025$ & $-0.022\pm0.009$ & $-0.096$ & \nodata \\
NGC\,6633 & 117 & 395 & $-0.122\pm0.011$ & $-0.084\pm0.110$ & $-0.038$ & \nodata \\
NGC\,2516 & 392 & 412 & $-0.102\pm0.008$ & $-0.076\pm0.025$ & $-0.025$ & 0.064 \\
NGC\,1901 & 60 & 416 & $-0.121\pm0.030$ & $-0.143\pm0.039$ & $+0.022$ & \nodata \\
NGC\,752 & 92 & 439 & $-0.058\pm0.014$ & $-0.040\pm0.010$ & $-0.018$ & \nodata \\
IC\,4756 & 188 & 474 & $-0.115\pm0.011$ & $-0.031\pm0.032$ & $-0.084$ & 0.069 \\
NGC\,3532 & 633 & 478 & $-0.070\pm0.007$ & $-0.027\pm0.060$ & $-0.043$ & 0.136 \\
NGC\,6281 & 266 & 531 & $-0.072\pm0.008$ & $-0.106\pm0.160$ & $+0.034$ & 0.042 \\
NGC\,2548 & 223 & 770 & $-0.044\pm0.012$ & $+0.013\pm0.057$ & $-0.057$ & 0.089 \\
NGC\,2168 & 754 & 871 & $-0.144\pm0.007$ & $-0.048\pm0.015$ & $-0.096$ & 0.003 \\
NGC\,2682 (M67) & 660 & 871 & $-0.045\pm0.008$ & $+0.010\pm0.030$ & $-0.055$ & 0.103 \\
NGC\,2447 & 377 & 1010 & $-0.182\pm0.008$ & $-0.080\pm0.010$ & $-0.102$ & 0.132 \\
NGC\,2437 & 1298 & 1706 & $-0.130\pm0.005$ & $+0.008\pm0.040$ & $-0.138$ & 0.008 \\
NGC\,188 & 630 & 1909 & $-0.053\pm0.005$ & $+0.090\pm0.020$ & $-0.143$ & 0.100 \\
NGC\,2158 & 754 & 4294 & $-0.364\pm0.008$ & $-0.210\pm0.020$ & $-0.154$ & 0.023 \\
\hline
\multicolumn{7}{l}{median $|\Delta| = 0.041$ dex,\quad mean $|\Delta| = 0.053$ dex} \\
\hline
\end{tabular}
\end{table*}

\section{Discussion}
\label{sec:discussion}

Across four independent tests i.e. a held-out LAMOST sample, two external surveys, the disk gradient, and open clusters, the model delivers $\feh$ estimates accurate to \SIrange{0.05}{0.07}{dex} which is roughly a five-fold improvement on GSP-Phot. The agreement with the \citet{Spina2022} cluster $\feh$ (\SI{0.041}{dex}) and the recovery of both the broken thin disk gradient and the open cluster gradient shows that predicted $\feh$ from our trained model can be used in population or other Galactic works.

We point to two residual systematics that deserve attention. First, the hump (Section~\ref{sec:hump}) appears to be a residual limitation of metallicity inference from XP-resolution $R\sim50$ and its minimum amplitude across our sample is $\sim\!\SI{0.07}{dex}$. So this sets a minimum uncertainty on the $\feh$ for any single star inferred from the model. Although, the hump may not badly affect the average $\feh$ of sample with wider temperature range, but it should be kept in mind when mixing populations with narrow colour distributions. Second, the cluster $\feh$ are underestimated for $d>\SI{1}{kpc}$.

We also note the inclusion of Galactic latitude as an input feature (Section~\ref{sec:gaia}). It carries little weight and leaves no trace in the latitude residuals (Figure~\ref{fig:resid_lat_plx}), but because it is a sky-position variable it could in principle let the model absorb part of the survey selection function. Anyone concerned with sky-dependent systematics should use this model with caution.

The model applies to AFGK stars within the colour, parallax, and quality ranges provided in Section~\ref{sec:xmatch}. It is not valid for O/B or very cool M stars which lie outside our training domain. We also point out that every input to the model, i.e. the XP coefficients, the GSP-Phot dust columns, and the astrometry, is a \emph{Gaia} DR3 product. One should, therefore, keep in mind that the estimates presented in this work are \emph{Gaia} DR3-based metallicities. We have not applied full dereddening, so the predictions in heavily extincted regions should also be treated with caution. The per-star uncertainty (Section~\ref{sec:uncertainty}) we get from the ensemble model is only a recommended reliability flag and not real uncertainty. We provide it so that the high-uncertainty predictions are not considered seriously, particularly in the metal-poor tail. When comparing with external surveys, the known $\lesssim0.1$--$0.2$~dex differences between spectroscopic scales should be taken into account.

\section{Conclusions}
\label{sec:conclusions}
We have estimated stellar metallicities from \emph{Gaia} DR3 XP data with a gradient-boosted decision-tree model trained on \num{1.20} million AFGK stars with LAMOST DR10 labels and \emph{Gaia}-only inputs. The main results are:

\begin{enumerate}
\item On the held-out LAMOST stars, the trained model has the MAE $=\SI{0.052}{dex}$ and $R^2=0.94$ with negligible bias, whereas for the same stars GSP-Phot has MAE $=\SI{0.24}{dex}$.

\item Our trained model transfers to external surveys with MAE $=\SI{0.066}{dex}$ (GALAH DR4) and \SI{0.068}{dex} (APOGEE DR17).

\item For \num{26} open clusters, it matches the high-resolution $\feh$ to a median \SI{0.041}{dex} which is smaller than an APOGEE trained XP model (\SI{0.067}{dex}) and GSP-Phot (\SI{0.248}{dex}).
\item When we apply our trained model to a kinematically selected sample of Bensby thin-disk stars drawn from the \num{4.85} million random stars, it recovers a broken thin disk gradient (inner $+0.119$, outer $-0.058\,\mathrm{dex\,kpc^{-1}}$, break near \SI{5.9}{kpc}). The 26-cluster open-cluster gradient from this work is $-0.066\, \mathrm{dex\,kpc^{-1}}$. Both values agree with high-resolution spectroscopy works.

\item We identify the intra-cluster colour ``hump'' possibly as an intrinsic limit of $R\sim50$ spectrophotometry which is also present in the APOGEE-trained model \citep{Andrae2023}.
\end{enumerate}

We provide the catalogue with $\feh$ and per-star uncertainties (derived from five model ensemble) for chemical studies of the Galactic disk.\footnote{The catalogue, the trained model, and the prediction scripts are publicly available on Zenodo at \href{https://doi.org/10.5281/zenodo.20645375}{doi.org/10.5281/zenodo.20645375}}

\emph{Gaia} DR4 is expected to deliver significantly improved XP spectra, and with them improved XP-based metallicities. Our model is trained on the DR3 XP coefficients and applying the same approach to DR4 would require a full re-training and re-validation on the new data. Even so, the metallicity estimates presented here, although based on \emph{Gaia} DR3, may remain competitive once DR4 becomes available.

\begin{acknowledgements}
We thank the anonymous referee for the constructive comments that improved this paper.
\\
R.S. acknowledges support from the National Science Centre, Poland, research grant 2022/47/I/ST9/02358
\\
This work has made use of data from the European Space Agency (ESA) mission {\it Gaia} (\url{https://www.cosmos.esa.int/gaia}), processed by the {\it Gaia} Data Processing and Analysis Consortium (DPAC, \url{https://www.cosmos.esa.int/web/gaia/dpac/consortium}). Funding for the DPAC has been provided by national institutions, in particular the institutions participating in the {\it Gaia} Multilateral Agreement.
\\
Guoshoujing Telescope (the Large Sky Area Multi-Object Fiber Spectroscopic Telescope LAMOST) is a National Major Scientific Project built by the Chinese Academy of Sciences. Funding for the project has been provided by the National Development and Reform Commission. LAMOST is operated and managed by the National Astronomical Observatories, Chinese Academy of Sciences.
\\
This work made use of the Fourth Data Release of the GALAH Survey (Buder et al. 2021). The GALAH Survey is based on data acquired through the Australian Astronomical Observatory. This paper includes data that has been provided by AAO Data Central (datacentral.org.au).
\\
\textit{Facilities:} \emph{Gaia}, LAMOST, GALAH (AAT), APOGEE (Sloan).
\end{acknowledgements}

\bibliographystyle{aa}
\bibliography{references}

\end{document}